\magnification=1200
\baselineskip=20 pt

\def\go{\tilde{g}_{w\tilde{\chi}\tilde{\gamma}}}
\def\gt{g_{\gamma\tilde{\chi}\tilde{\chi}}}
\def\gth{g_{\gamma ww}}
\def\m{\tilde{m}}
\def\M{\tilde{M}}

\def\zw{Z_w}
\def\zg{Z_{\gamma}}
\def\zph{Z_{\tilde{\gamma}}}
\def\zcgw{Z_{\tilde{\chi}\tilde{\gamma}}}
\def\zc{Z_{\tilde{\chi}}}
\centerline{\bf Violation of SUSY equivalence in triple gauge boson }

\centerline{\bf and gaugino couplings }

\vskip 1 true in

\centerline{\bf Uma Mahanta}

\centerline{\bf Mehta Research Institute}

\centerline{\bf Chhatnag Road, Jhusi}

\centerline{\bf Allahabad-211019, India}

\vskip .4 true in

\centerline{\bf Abstract}

Supersymmetry implies that the trilinear couplings between gauge bosons
and gauginos are equal to all orders. However if SUSY is broken and
some of the superpartners have large SUSY breaking masses then they can
induce non-decoupling deviations in this relation through radiative
corrections. In this work we show that the deviation in the ratio
${\go \over \gth}$ from unity can be around 2.3\% (3.3\%) in heavy QCD
models (2-1 models) for a point in the gaugino region. Precision
measurement of triple gaugino couplings at future high energy colliders 
can therefore probe such violations and the existence of heavy
superparticles in the multi Tev region which are otherwise inaccessible.

\vfill\eject

\centerline{\bf Introduction}

In recent years SUSY has emerged as a leading candidate for the origin of
electroweak symmetry breaking. If SUSY is to provide a cure for the
heirarchy problem and the quadratic divergence problem then the superparticles
must lie between  100 Gev and a few Tev range otherwise various fine tunings
[1] are required to reconcile it with low energy constraints.
 Most of the future
colliders are therefore being designed with a view to produce these 
sparticles which would lead to the discovery of SUSY. After their discovery
we need to verify whether these particles are really the SUSY partners
of ordinary particles. SUSY implies that the couplings of superparticles
be equal to the corresponding couplings of ordinary particles. In particular
if $g_i$ are the SM gauge couplins and $\tilde {g}_i$ are the corresponding
superpartner couplings, then exact SUSY implies that $g_i = \tilde{g}_i$
to all orders. However if SUSY is broken and some of the superpartners
have large SUSY breaking mass splittings then they can induce violations
to the above relation through radiative corrections. These violations
are similar to the violations in custodial $SU(2)$ symmetry [2] due to a
split $SU(2)$ multiplet. In fact if some of the sparticles have masses
between 1-10 Tev then they will be inaccessible at the planned high energy
colliders. Although they decoulpe from most low energy processes their
contribution to the difference ${\tilde{g}_i\over g_i}-1$ grows 
logarithmically with the heavy superpartner mass scale. Hence they induce
a non-decoupling effect in low energy processes involving light
sparticles and ordinary particles. Precision measurements of the
couplings of light sparticles at future high energy colliders can therefore
be used to probe such deviations from exact supersymmetric relations
 and also the heavy superpartner mass scale. 

\centerline{\bf Violation of SUSY equivalence between gauge}

\centerline{\bf and gaugino couplings}

In Ref. 3 Cheng et al considered the deviation of the gaugino couplings
to $f-\tilde{f}$ from the corresponding SM gauge coupling. They chose to
parameterize these deviations in terms of three superoblique parametrs
$\tilde{U}_i={\tilde{g}_i\over g_i}-1$ where the subscript i denotes the 
relevant SM gauge group. These authors considered precision measurements in
chargino pair production at $e^+e^-$ collider and selectron pair production at
$e^+e^-$ and $e^-e^-$ colliders. Since the cross-section in these processes
 measure the gaugino couplings to $f-\tilde{f}$ pair, the deviations in these
couplings from the corresponding SM gauge couplings are the relevant
ones to consider. They showed that $\tilde{U}_1\approx .35\%(.29\%)\times
\ln R$ and $\tilde{U}_2\approx .71\%(.80\%)\times \ln R$ for 2-1 models
(heavy QCD models). Here $R={\M \over \m}$ is the ratio of heavy to light
sparticle mass scales. The light sparticle mass scale corresponding to
electroweak gauginos is typically of the order of 100 Gev. The heavy 
superpartner mass scale $\M$ is usually around 1 Tev for heavy QCD
models like gauge mediated SUSY breaking models [4] where strongly interacting
sparticles get large masses through QCD ineractions.
 There is another class
of models known as 2-1 models [5] where some of the sparticles can be very
heavy. In these models the sfermions belonging to the first two generations
lie between 10-50 Tev but the third generation sfermions are near the weak 
scale. The estimates for $\tilde {U}_1$ and $\tilde{U}_2$ considered by
Cheng et al therefore lie between 1.61\%-3.27\% (.67\%-1.84\%) for 
2-1 models (heavy QCD models)
 depending upon  the relevant EW gauge group.

It would be interesting to know what is the deviation of gaugino coupling
from the gauge coupling in the pure gauge sector namely the triple
gauge boson couplings. The supersymmetric version of Slavnov Taylor
identity implies that the gaugino coupling 
$\tilde{g}_{\tilde{\chi}f\tilde{f}}$    to 
$f-\tilde {f}$ is equal to $\go$ to all orders if SUSY is unbroken.
However if SUSY is broken then the Slavnov-Taylor identity in the sparticle
 sector also breaks down. Therefore it is not apriori clear if the violation
of SUSY equivalence in $\go$ would be of the same
order as that in $\tilde{g}_{\tilde{\chi}f\tilde{f}}$. Further in unbroken
 SUSY the
 Slavnov-Taylor identity only equates the leading logarithm corrections
in $\go$ and $\tilde{g}_{\tilde{\chi}f\tilde{f}}$, but not the finite terms
which are quite important in our case since the leading log terms are not 
very large.
 In this work we shall consider
the violation of SUSY equivalence between trilinear couplings of gauge
bosons and gauginos. We shall show that in heavy QCD models where $R\approx 10$
the correction is of the order of 2.3\%. On the other hand in 2-1 models
where $R\approx 100$
the correction can be as large as 3.3\%. Therefore if the trilinear 
gaugino Yukawa coupling
can be measured with an accuracy of about 1\% from chargino decay
at future high energy colliders then not only
we can probe such deviations but also the heavy sparticle mass scale.

\centerline{\bf Specification of the model and assumptions}

We shall evaluate the radiative corrections in two distinct scenarios. Under
the first scenario we shall consider heavy QCD models where all strongly
interacting sparticles are very heavy and lie in the few Tev range. The 
sparticles which have only EW quantum numbers will be assumed to be light
and in the 100 Gev range. Under the second scenario we shall consider the
2-1 models which were proposed to solve the SUSY flavor problem and CP problem
without requiring degeneracy, alignment or small CP phases in the squark
mass matrices.
 In these 
models all sparticles belonging to the first two generations are very
 heavy and they decouple from  the remaining sparticles comprising the squarks 
and sleptons of the third generation and gauginos. Bounds from flavor
 changing neutral currents imply that the heavy superpartner mass scale
 in 2-1 models can be as large as 10-50 Tev [5]. 

We shall evaluate the relevant one loop diagrams using the $\bar{MS}$
scheme. More accurately we should use the SUSY preserving $\bar{DR}$
scheme. But since the loop diagrams in our case involves only fermions and 
sfermions the $\bar{DR}$ scheme give the same ressult as the  $\bar{MS}$
scheme. We shall set the arbitrary renormalization mass scale
$\mu$ equal to the
light superpartner mass scale $\m$. This is the superpartner mass scale
that will be accessible at planned future colliders. To simplify our
calculations we shall assume that the lightest chargino is dominantly
a gaugino and the lightest neutralino is a photino. If the lightest chargino
and neutralino have significant higgsino components then there will be 
additional contributions from the Yukawa couplings of higgsino to top quark
or top squark that cannot be neglected. However our results for 2-1 models
 where the third generation sfermions are light will be almost
unaffected as we move
 from the gaugino dominated region to the higgsino region.

\vfill\eject
\centerline{\bf Radiative corrections in heavy QCD models}

In this section we shall calculate the radiative corrections to the couplings 
$\go$, $\gt$ and $\gth$ in heavy QCD models. Unbroken SUSY implies that
$\go =\gt =\gth$ to all orders. However if SUSY is broken then there will be
small deviations from it induced by heavy superpartners. We shall assume that 
the external momenta of any loop diagram satisfy the condition
 $p_i^2\approx \m ^2 \ll \M ^2$. In evaluating the renormalization constants
 from the loop diagrams we shall retain both heavy and light particles. 
However loop diagrams that do not contain any heavy particles do not give
 rise to large logarithms of the form $\ln {\M ^2\over \mu ^2}$ and 
can be neglected. In our case the leading log terms are not very large.
Therefore besides the leading logs we shall keep all the finite terms that do
 not vanish in the limit ${\m ^2\over \M ^2}\rightarrow 0$. 

The renormalization of the coupling constant $\go $ is given by
${\go (\M )\over \go (\mu )}= {Z_{\tilde {\chi}\tilde {\gamma}W}\over
\sqrt {Z}_{\tilde {\chi}}\sqrt {Z}_{\tilde {\gamma}}\sqrt {Z}_w}$
where $Z_{\tilde {\chi}\tilde {\gamma}W}$ is the vertex renormalization
constant. $\sqrt {Z}_{\tilde {\chi}}$, $\sqrt {Z}_{\tilde {\gamma}}$
$\sqrt {Z}_w$ are the wavefunction renormalization constants associated
with chargino, photino and W boson fields respectively.
 Due to the Ward identity
associated with the unbroken $U(1)_q$ gauge symmetry the renormalization of
$\gt$ and $\gth$ are given by ${\gt (\M )\over \gt (\mu )}={\gth (\M )\over
\gth (\mu )}={1\over \sqrt {Z}_{\gamma}}$.
In heavy QCD models $(N_f=N_c=3)$ we get

$$Z_{\tilde {\chi}\tilde {\gamma}W}=1-{g^2\over 32 \pi^2}N_fN_c(\ln R^2-
{1\over 2}).\eqno(1)$$

$${1\over \sqrt {Z}_{\tilde {\chi}}}=1+{g^2\over 64 \pi^2}N_fN_c(\ln R^2-
{1\over 2}).\eqno(2)$$

$${1\over\sqrt {Z}_{\tilde {\gamma}}}=1+{e^2\over 32 \pi^2}{5\over 9}
N_fN_c(\ln R^2-{1\over 2}).\eqno(3)$$

$${1\over \sqrt {Z}_w}=1+{g^2\over 64 \pi^2}N_fN_c\ln R^2.\eqno(4)$$

and

$${1\over \sqrt {Z}_{\gamma}}=1+{e^2\over 32 \pi^2}{5\over 9}
{N_fN_c\over 3}\ln R^2.\eqno(5)$$

Here $R={\M \over\mu}$. We expect SUSY to be restored  above the heavy
superpartner mass scale $\M$.
 Using the boundary condition  $\go (\M ) = \gt (\M ) =\gth (\M )$
we get 

$$\eqalignno{{\go (\mu )\over \gt (\mu )}&={\go (\mu)\over \gth (\mu )}\cr
&=[1+{g^2\over 64\pi^2}N_fN_c  \{(1
-{10\over 9}  s_w^2)(\ln R^2-{1\over 2})\cr
&-{1\over 3}\ln R^2 \}+{e^2\over 96\pi^2}{5\over 9}N_fN_c\ln R^2]\cr
&\approx 1+.023\ .&(6)\cr}$$

Note that when SUSY is broken and squarks get a large mass, quark loops
do not contribute to large logarithms and hence their contribution to
$Z_w$ and $Z_{\gamma}$ can be neglected. However in the SUSY limit the
contribution of quark loops has to be added. When that is done we get

$${1\over\sqrt {\zw }}=1+{g^2\over 64\pi^2}N_fN_c\ln R^2={1\over\sqrt
{Z_{\tilde{\chi}}}}.\eqno(7)$$

$${1\over \sqrt{\zg }}=1+{e^2\over 32\pi^2}{5\over 9}N_fN_c\ln R^2=
{1\over\sqrt{\zph }}.\eqno(8)$$

The renormalization of $\go $ in the SUSY limit is therefore given by

$${\go (\M )\over\go (\mu )}={1\over \sqrt{\zg}}={1\over \sqrt{\zph}}
={\gth (\M )\over \gth (\mu )}.\eqno(9)$$

This is in accord with the SUSY version of of Slavnov-Taylor identity
which demands that in the SUSY limit $\go $ should renormalize exactly
like $\gth$. However if SUSY is broken then Slavnov Taylor
 identity breaks down for $\go $.
The Wavefunction renormalization constants of W and $\chi$ no longer 
cancel the vertex renormalization constant $Z_{\tilde{\chi}\tilde{\gamma}W}$
 completely. As a result $\go$ receives its dominant renormalization 
from asymptotically free $SU(2)_w$  interactions.

\centerline{\bf Radiative corrections in 2-1 models}

In this section we shall present the results for radiative corrections
to $\go$, $\gt$ and $\gth$ in 2-1 models. Evaluating the one loop diagrams
 we get

$$\zcgw =1-{g^2\over 4\pi^2}(\ln R^2-{1\over 2}).\eqno(10)$$

$${1\over \sqrt {\zc}} =1+{g^2\over 8\pi^2}(\ln R^2-{1\over 2}).\eqno(11)$$

$${1\over \sqrt {\zph}} =1+{e^2\over 6\pi^2}(\ln R^2-{1\over 2}).\eqno(12)$$

$${1\over \sqrt {\zw}} =1+{g^2\over 24\pi^2}\ln R^2.\eqno(13)$$

and

$${1\over \sqrt {\zg}} =1+{e^2\over 18\pi^2}\ln R^2.\eqno(14)$$

The effect of supersymmetry breaking on the trilinear gauge 
 couplings in 2-1 models is 
given by

$$\eqalignno{{\go (\mu )\over \gt (\mu )}&={\go (\mu )\over \gth (\mu )}\cr
&=[1+{g^2\over 4\pi^2}\{{1\over 3}(1-{4\over 3}s_w^2)\ln R^2
-{1\over 2}({1\over 2}-{2\over 3}s_w^2)\}]\cr
&\approx 1+.033 .&(15)}$$

The same results as above can also be obtained by decoupling the heavy fields
at $\M$ and evolving the couplings down to $\mu$
according to the low energy theory
of light degrees of freedom.
Comparing our result with that of Cheng et al we can conclude that the
 violation of SUSY equivalence in trilinear gaugino coupling is of the same
order as that in $\tilde{g}_{\tilde{\chi}f\tilde{f}}$. Hence the violation of
Slavnov-Taylor identity that relates the renormalization of $\go$ to that of
 $\tilde{g}_{\tilde{\chi}f\tilde{f}}$
 is either very small or its
effect on the quantity under study is negligible.

\centerline{\bf Experimental prospects}

The results obtained in the previous sections imply that
if the gaugino coupling $\go$ can be measured with an accuracy of around 1\%
then we can probe such violations in supersymmetric relations and also the
heavy superpartner mass scale. If the decay $\tilde{\chi}\rightarrow
\tilde{\gamma}w$ is kinematically allowed then it would provide the best 
source for measuring the coupling $\go$. In fact in some models like the
gauge mediated SUSY breaking the LH slepton doublet is heavier than the
lightest chargino. This makes  $\tilde{\chi}\rightarrow
\tilde{\gamma}w$ the dominant decay mode with an appreciable branching ratio.
At a 500 Gev $e^+e^-$ collider the chargino pair production cross-section for
LH incoming $e^-$ beam, $M_{\tilde{\chi}}= 170$ Gev and
 $m_{\tilde{\nu}_e}=200$ Gev
 is about 600 fb. With a design luminosity of 50 fb$^{-1}$/yr, 60000
charginos will be produced. If the decay  $\tilde{\chi}\rightarrow
\tilde{\gamma}w$ has a branching fraction of .8 then 48000 charginos will
decay in this mode. This would yield a statistical error of .46\% in the
chargino decay width via this mode
and .23\% in the coupling $\go$. However we also need to 
calculate the theoretical systematic error arising out of uncertainties in
 $M_{\tilde{\chi}}$, $M_{\tilde{\gamma}}$ and $m_{\tilde{\nu}_e}$ which 
would affect the determination of $\go$. Further 
all the relevant backgrounds have to be computed and subtracted from the
 signal to reduce the total uncertainty in the decay width. With a bit of 
optimism it may be fair to say that although
it may not be possible to measure $\go$ with the same degree of accuracy
as $\tilde{g}_{\tilde{\gamma}\tilde{e}e}$ or
 $\tilde{g}_{\tilde{\chi}\tilde{\nu}e}$ the prospects for measuring it
 with a somewhat reduced degree of accuracy seem quite bright.

\centerline{\bf References}

\item{1.} J. Ellis, K. Enqvist, D. Nanopoulos and F. Zwirner, Mod. Phys.
Lett. A 1, 57, (1986); R. Barbieri and G. F. Giudice Nucl. Phys. B 306, 63 
(1988); G. W. Anderson  and D. J. Castano, Phys. Lett. B 347,
300 (1995); Phys. Rev. D 52, 1693 (1995); 53, 2403 (1996).

\item{2.} P. Sikivie, L. Susskind, M. Voloshin and V. Zakharov, Nucl. Phys.
B 173, 189 (1980).

\item{3.} H. -C. Cheng, J. L. Feng and N. Polonsky, Phys. Rev. D 56, 6875
(1997); H. -C. Cheng, J. L. Feng and N. Polonsky, Phys. Rev. D 57, 152
(1998).

\item{4.} M. Dine and A. Nelson, Phys. Rev. D 48, 1277(1993); M. Dine,
A. Nelson and Y. Shirman, Phys. Rev. D 51, 1362 (1995); M. Dine, Y. Nir
and Y. Shirman Phys. Rev. D 53, 2685 (1996).

\item{5.} A. G. Cohen, D. B. Kaplan and A. E. Nelson, Phys. Lett. B 388,
588 (1996); A. G. Cohen, D. B. Kaplan, F. Lepintre and A. E. Nelson, Phys.
Rev. Lett. 78, 2300 (1997); A. E. Nelson and D. Wright Phys. Rev. D 56,
1598 (1997).

\end